\documentclass[aps,showpacs,twocolumn,floatfix,pra]{revtex4}
\usepackage{graphicx}
\usepackage{amsfonts}
\usepackage{amsmath}
\usepackage{amssymb}

\begin{document}
\title{Levy distribution in many-particle quantum systems}
\author{A. V. Ponomarev}
\author{S. Denisov}
\author{P. H\"{a}nggi}
\affiliation{Institute of Physics, University of Augsburg,
Universit\"{a}tstr.~1, D-86159 Augsburg}
\date{\today}
\begin{abstract}
Levy distribution, previously used to describe complex behavior
of classical systems, is shown to characterize that of quantum
many-body systems. Using two complimentary approaches, the canonical
and grand-canonical formalisms, we discovered that the momentum profile
of a Tonks-Girardeau gas, -- a one-dimensional gas of $N$ impenetrable
(hard-core) bosons, harmonically confined on a lattice at finite temperatures, 
obeys Levy distribution. Finally, we extend our analysis to different 
confinement setups and demonstrate that the tunable Levy distribution 
properly reproduces momentum profiles in experimentally accessible regions. 
Our finding allows for calibration of complex many-body quantum
states by using a unique scaling exponent.
\end{abstract}


\maketitle

\section{Introduction}

Since the first observations of quantum collective phenomena, the quantum
systems with strongly interacting constituents have become of paramount
interest in condense matter community \cite{thoulfetter}.
A new wave of activity in the area of quantum many-body systems has been
burgeoning with the advent of laser cooling techniques \cite{metcalf}.
Many of those quantum models, which were thought to be theoretical
abstractions, have since been implemented with cold atoms \cite{ober,zwerger}.
One of such models, a one-dimensional system of hard-core bosons --
a Tonks-Girardeau (TG) gas \cite{tonk} -- has been thoroughly
probed in recent experiments \cite{bloch, kinoshita1}.

While the density profile and the energy spectrum of hard-core bosons
resemble that of non-interacting fermions, its momentum distribution
(MD) exhibits distinct features. The ground state of a homogeneous TG gas
is known to possess an infrared divergence in the thermodynamic
limit, $N \rightarrow \infty$, $n(p) \propto p^{-1/2}$
\cite{olshani}, which, however, vanishes upon addition of a harmonic
confinement \cite{papen}. So far, there are no analytic results on the
infrared behavior of the finite-temperature MD for a finite number of
bosons in a harmonic trap. In a sole harmonic confinement, the
ground-state MD decays as $n(p)\propto p^{-4}$ at the high momentum
regime \cite{minguzzi}. Yet in the presence of an optical lattice,
which sets an upper momentum scale given by the recoil momentum,
$\hbar k_L$, where $k_L$ is the wave vector of the laser beam,
this region cannot be resolved in present state-of-the-art experiments.

The finite-temperature MDs of $N \leq 20$ strongly repulsive
bosons, confined on a one-dimensional optical lattice and an
additional harmonic trap (a so called ``1d tube''), have been
measured experimentally \cite{bloch}. These measurements revealed
that in the intermediate region, $0.3 \leq |p/\hbar k_L| \leq 1$, a
momentum profile can be approximated by a power-law,
$n(p) \propto p^{-\gamma}$. The exponent $\gamma$  depends on
temperature, density, and strength of the atom-atom interactions.
Results of numerical Monte-Carlo simulations have corroborated the
experimental finding \cite{pollet}.

What kind of momentum distribution emerges in a system of
strongly repulsive bosons? In the present work, we attest that
{\it Levy distribution} describe the MD of $N$ thermalized
hard-core bosons in various one-dimensional confinements,
in particular, within a single 1d tube as well as within array of
1d tubes probed experimentally \cite{bloch, kinoshita1}.
Levy statistics \cite{Levy} are known to describe classical chaotic
transport \cite{Klafter}, processes of subrecoil laser
cooling \cite{sub}, fluctuations of stock market indices
\cite{econ}, time series of single molecule blinking events
\cite{barkai}, or bursting activity of small neuronal networks
\cite{network}. The appearance of Levy distribution in a system output
is a strong indicator of a long-range correlation ``skeleton'' which
conducts system intrinsic dynamics \cite{network, heart}.
However the Levy distribution has at no time emerged in the context
of {\it many-particle quantum} systems before. The great advantage of
the Levy-based analysis is its capability of calibration of the TG
in different quantum regimes by a unique scaling exponent $\alpha$.

The paper is organized as follows. In Sec.~II, we introduce the lattice
TG model and employ two complimentary approaches to study its
finite-temperature properties. In Sec.~III, we fit exact results of
the calculation within the grand-canonical formalism \cite{grandcanon}
by the Levy distribution. To provide an insight into experimental situation
\cite{bloch}, we devoted Sec.~IV to the analysis of momentum profiles
averaged over the array of 1d tubes. We proceed with the results obtained
within the canonical formalism \cite{canon} and conclude the section with
the analysis of the experimental data from the Ref. \cite{bloch}. In Sec.~V,
we demonstrate the universality of the Levy spline approximation by addressing
the MD of a TG gas in one-dimensional confinements of various geometry. 
We elaborate on the case of a TG gas in a sole harmonic confinement, 
in a box, and in a sole optical lattice with impenetrable (hard wall) 
boundaries. Finally, in Sec.~VI, we summarize our results. Some of important 
technical details are deferred to the Appendixes A and B.

\section{Tonks-Girardeau gas at finite temperatures on a 1d optical 
lattice}

A Bose gas confined in a deep optical lattice is well described
by the Bose-Hubbard Hamiltonian \cite{BHH}
\begin{equation}
\label{eq:hamiltonian} H = -
J\sum_l(b_l^{\dagger}b_{l+1}+h.c.)+\nu\sum_l l^2 n_l +\frac{U}{2}
\sum_l n_l(n_l-1),
\end{equation}
where $n_l = b_l^{\dagger}b_l$ is the particle number operator on lattice 
site $l$, $J$ is the hopping strength, the parabolicity $\nu=M\omega_0^2d^2/2$ 
is the amplitude of the external harmonic potential with the trapping frequency 
$\omega_0$, $M$ is the mass of the atom, and $d$ is the lattice constant. The 
last term in (\ref{eq:hamiltonian}) describes the on-site atom-atom interactions. 
Here we are interested in the TG regime, where the strength of the repulsive atom-atom 
interaction considerably exceeds the kinetic (hopping) energy, i.e., $U/J \rightarrow \infty $,  
\cite{tonk}. Therefore, the interaction term can be substituted by the condition 
that two bosons cannot occupy the same lattice site \cite{tonk}.

\begin{figure}[t]
\center
\includegraphics[width=0.45\textwidth]{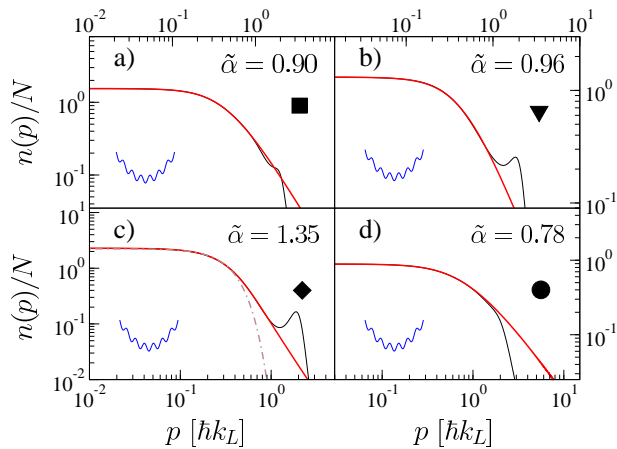}
\caption {(Color online) Normalized momentum density profiles (MD) for $N = 40 \textrm{(a)};~ 30\textrm{(b)};~ 10\textrm{(c)};~90\textrm{(d)}$ hard-core
bosons in a 1d tube, thin (black) lines, at four different sets of
parameters are compared to Levy distributions, thick (red) lines. Dashed-doted line (c)
corresponds to a Gaussian approximation. The parameters $(N/n_s,k_BT/J)$
are indicated by symbols $\blacklozenge$, $\blacktriangledown$, $\blacksquare$,
$\mathbf{\bullet}$ in Fig. 2. The thin solid (blue) icon indicates the confinement 
geometry.} \label{fig1a}
\end{figure}

The momentum distribution can be obtained from the reduced single-particle
density matrix $\rho_{n,l}$, reading
\begin{equation}
\label{eq:moment}
n(p)=|\Phi(p)|^2\sum_{n,l}e^{-ip(n-l)}\rho_{n,l},
\end{equation}
where the momentum $p$ is written in units of recoil momentum $\hbar
k_L = \pi\hbar/d$, and the envelope $\Phi(p)$ is the Fourier transform of
the Wannier function. Note that the latter, as well as the hopping
$J$, is solely defined by the lattice depth, $V_0$, measured in units
of recoil energy, $E_R=(\hbar k_L)^2/2M$.

To find the reduced single-particle density matrix, we employ here the
grand-canonical \cite{grandcanon} and the canonical formalism
\cite{canon} (See Appendixes A, B). The first is relevant
for a system being in contact with a thermal cloud at constant temperature
$T$, while the second describes an isolated many-particle system.

The difference between the momentum distributions obtained
within the grand-canonical and the canonical descriptions is mediated
by the number of particles and becomes negligible for $N\gtrsim10$
\cite{grandcanon}. In addition, at finite temperatures $k_BT\gtrsim 0.1J$,
systems with different number of particles, but the same densities $N/n_s$,
possess the same momentum profiles \cite{grandcanon}. Here, $n_s = 8(J\nu)^{1/2}/\pi$
is the number of single-particle eigenstates with non-zero population
at the trap center \cite{pendulum}. The latter also yields the critical
number of bosons in a 1d tube required to form the Mott-insulator in the
trap center at zero temperature.

The typical examples of the MD obtained here within the grand-canonical
formalism are presented in Fig. 1 for $N \geq 10$, where the results of both
the grand-canonical and canonical descriptions practically identical.

\section{Levy spline approximation}

In view of the strong non-Gaussian behavior of the function $n(p)$
(see Fig. 1c) and an apparent power-law intermediate asymptotics
\cite{bloch, pollet}, it is tempting to compare the MD of a TG gas
with symmetric Levy distribution \cite{Levy}. The latter, $L_{\alpha}(p)$,
is a natural generalization of the Gaussian distribution, and it
is defined by the Fourier transform of its characteristic function; i.e.,
\begin{equation}
\label{eq:Levy} L_{\alpha}(p) =
\frac{1}{2\pi}\int_{-\infty}^{\infty} \varphi(\xi)e^{i p \xi}
d\xi,~~~\varphi(\xi)=exp(-C_{\alpha}|\xi|^{\alpha}) \,,
\end{equation}
where the exponent $0 < \alpha \leq 2$, and $C_{\alpha}> 0$ is some
constant \cite{Levy}. For $\alpha=2$ the distribution is the Gaussian.
The value of $\alpha$ is invariant under the scaling transformation
$p \rightarrow const \cdot p$, which can be used for fitting.

The Levy distribution exhibits a power-law asymptotics, $L_{\alpha}(p)
\propto p^{-(\alpha+1)}$, as $|p| \rightarrow \infty$. These
``heavy'' tails  cause the variance of Levy distributions to diverge
 for all $\alpha < 2$. However, this asymptotic limit is not
relevant for our objective: the MD for the system (1) is bounded by
the width of $|\Phi(p)|$, $(V_0/E_R)^{1/4} \hbar k_L$
\cite{pollet}, so that the MD variance remains finite.

\begin{figure}[t]
\center
\includegraphics[width=0.45\textwidth]{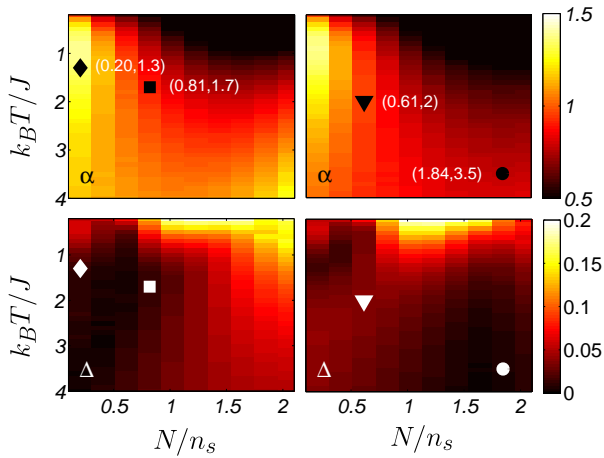}
\caption {(Color online)
The scaling exponent of Levy distribution, $\alpha$ (top), and the
corresponding root mean square deviation from the exact momentum
density profile, $\Delta$ (bottom), of TG gas as a function of
temperature and particle density  for two different amplitudes of
the optical lattice,  $V_0=4.6E_R$ (left column), and $V_0=9.3E_R$
(right column), see text for more details. The amplitude of harmonic 
confinement is chosen as $\nu=2\cdot 10^{-4} E_R$.} \label{fig1}
\end{figure}

The fitting of the calculated MD $n(p)$ by the Levy distribution
$L_\alpha(p)$ was automatized and performed numerically by the
global minimum search of the root mean square deviation, $\Delta =
\left\{\sum_n[n(p_n)-L_\alpha(p_n)]^2\right\}^{1/2}/n(0)$, in a 2d
parametric space $\{\alpha, C_\alpha\}$ (\ref{eq:Levy}). For the results
shown on Figs.~1,2, we took $50$ spline points equally spaced on the
interval $p\in(0,\hbar k_L)$. The initial area, ${\alpha\in(0,2),C_\alpha\in(0,30)}$,
covered by $10\times10$ grid was iteratively converged to the global
minimum by decreasing the grid step in $\alpha$ and $C_\alpha$ until the
desired relative accuracy (fixed to $1\%$ in all figures) was reached.

In Fig. 2, we show the dependence of scaling exponent $\alpha$ (top)
on the scaled temperature $k_BT/J$ and the scaled particle density, $N/n_s$
(obtained for $N=10,...,100$, and $n_s=48.78$), together with
the associated mean square deviation $\Delta$ (bottom), at
relatively low and high amplitude of the optical lattice. These
diagrams (see also Fig.~1, for concrete examples of the MD and their
Levy-spline approximations) constitute the first main result of this
work, namely, the convergence of the MD of a TG gas towards the Levy
distribution with increasing the temperature.

Despite of infinitely strong on-site repulsive interactions, the
systematic increase of the Levy exponent $\alpha$ with
temperature, cf. Fig. 2 (top), is consistent with the high-temperature
limit where the ideal Bose gas obeys the classical Boltzmann-Maxwell
statistics with a Gaussian MD, i.e., $\alpha=2$.

\section{Averaging over the array of 1d optical lattices}

The single 1d tube realization, using a fixed temperature $T$ and
a fixed particle number $N$, is not directly accessible with the present
state-of-art experiments. The averaging over an array of tubes done in Ref.
\cite{bloch} can be understood as an averaging over many realizations,
with different parameters, $T$ and $N$.

Namely, the experimental setup \cite{bloch} produces an array of independent
1d tubes, with different numbers of particles, $N_i$. The probability of
having a tube with $N$ particles is given by \cite{bloch}
\begin{equation}
\label{eq:averag}
\varrho(N)= \frac{2}{3} \frac{1}{N^{2/3}_c N^{1/3}},~~N \leq N_c,
\end{equation}
where the number of particles in a central tube $N_c$ is a unique parameter.
Assuming the same initial temperature in all the tubes for a shallow 1d
lattice potential, $V_{\rm in}$, the temperatures at the experimentally adjusted
lattice depth, $V_0$, can be obtained by using the conservation of entropy
in each tube during the subsequent adiabatic increase of the lattice depth from
$V_{\rm in}$ to $V_0$. Therefore, tubes with the different number of particles, $N_i$,
acquire different final temperatures, $T_i$, at $V_0>V_{\rm in}$ \cite{bloch}.

\begin{figure}[t]
\center
\includegraphics[width=0.45\textwidth]{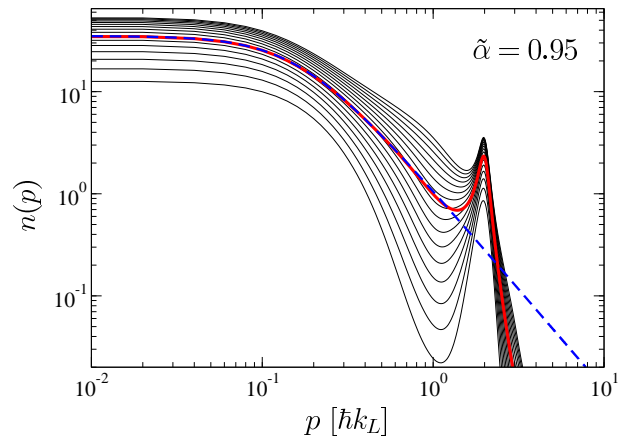}
\caption {(Color online) Unnormalized momentum
distributions (MD) for $N=3,...,18$ hard-core bosons, thin (black) lines,
obtained within the canonical approach (\ref{eq:cdmatrix}) and the
MD averaged over the array of 1d tubes according to the experimental
procedure used in Ref. \cite{bloch}, thick (red) line, with its
best Levy-spline, thick (blue) dashed line.}
\label{fig2}
\end{figure}

We implemented this averaging procedure with $N_{c}=18$, for a set of
individual MDs pre-calculated within the canonical formalism. The result
for $k_BT=0.5J$, $V_0=V_{\rm in}=4.6E_R$ and $\nu = 8\cdot 10^{-4} E_R$ is
depicted in Fig. 3.

Surprisingly enough, the averaged MD can be perfectly approximated by
the Levy distribution with an ``average'' scaling exponent, $\tilde{\alpha}$;
this is so despite the sizable dispersion of $\alpha$-values appearing in
the different momentum profiles.

Therefore, the averaged momentum profile appears as a superposition of several
different profiles. It is known that using a proper weight function, $f(\beta)$,
one can construct a Levy distribution from a parameterized set of Gaussian
distributions of different dispersion, $L_{\alpha}(p)=\int_{-\infty}^{\infty} G(p,\beta)f(\beta)d\beta$
\cite{superstat}. Yet, to the best of our knowledge, there are no results
concerning the superposition of many different Levy distribution functions with
\textit{different exponents} $\alpha$.

A next objective is the comparison of our scheme with the experimental
data from Ref.~\cite{bloch}, see Fig.~4. As one can deduce the Levy
distribution yields an excellent approximation for the MD of the experimental
system although the latter does not map precisely to a TG gas, but rather
corresponds to a set of $N$ soft-core bosons with
$U/J < \infty$ (\ref{eq:hamiltonian}) \cite{pollet}. In the experiment,
one used a power-law fit, $n(p) \propto p^{-\gamma_{exp}}$, on an intermediate
range ($0.3 \leq |p/\hbar k_L| \leq 1$) \cite{bloch}, with
$\gamma_{\rm exp} -1 = 1.2; ~0.9; ~0.4; ~0.2$ for the data shown on Figs.~4(a-d),
respectively.  We emphasize, however, that in the experimentally accessible region,
$|p|\lesssim \hbar k_L$, the power law behavior with the exponent
$\gamma_{\rm exp}-1=\tilde{\alpha}$ of the Levy distribution is not yet valid,
but is assumed only for much larger momentum values. Therefore, the theoretical
estimates $\tilde{\alpha}$ in Fig.~4 exceed those intermediate range
power-law fit-values, i.e. one consistently finds that $\tilde{\alpha} > \gamma_{\rm exp}-1$.

\begin{figure}[t]
\center
\includegraphics[width=0.45\textwidth]{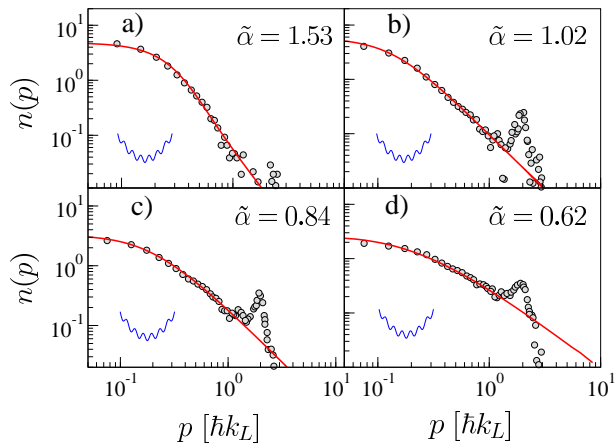}
\caption {(Color online) The experimental data
for the MD of the 1d quantum gases for different axial lattice depths
\cite{bloch}, grey dots; Levy distributions with the scaling
exponent $\tilde\alpha$, solid (red) lines. The thin solid (blue) icon 
indicates the confinement geometry.} \label{fig3}
\end{figure}

\section{Universality of the Levy-spline for a Tonks-Girardeau
gas in 1d confinement potentials}

\begin{figure}[t]
\center
\includegraphics[width=0.45\textwidth]{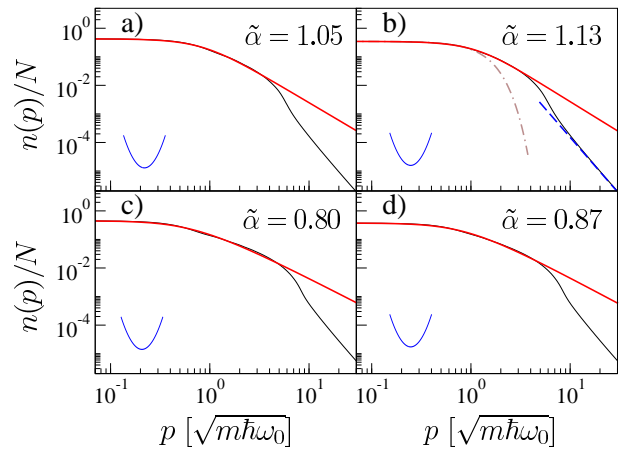}
\caption {(Color online) Normalized momentum density profiles (MD)  for $N = 5 \textrm{(a,b)};~ 10\textrm{(c,d)}$ hard-core
bosons in a sole 1d harmonic trap, thin (black) lines, at the two different temperatures,
$k_BT/\hbar\omega_0=1$ \textrm{(a,c)}, $k_BT/\hbar\omega_0=1.5$ \textrm{(b,d)}
are compared to Levy distributions, thick (red) lines. Dashed line (c) underlines well known
universal power-law asymptotic behavior of the Tonks-Girardeau gas momentum
distribution at large momenta, $p^{-4}$, while dashed-dotted line illustrates failure of a Gaussian
approximation. The thin solid (blue) icon indicates the confinement geometry.}
\label{fig4}
\end{figure}

\begin{figure}[t]
\center
\includegraphics[width=0.45\textwidth]{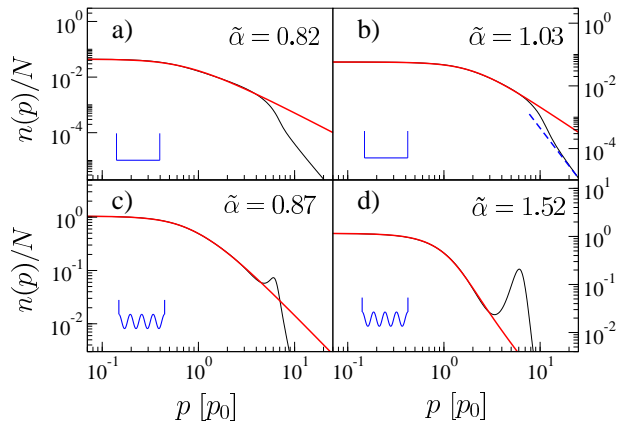}
\caption {(Color online) Normalized momentum density profiles (MD) 
for $N = 10 \textrm{(a,d)};~ 5 \textrm{(b)};~50\textrm{(c)}$ hard-core
bosons in a 1d box (a,b) of size $L_{\rm box}$ and on a lattice (c,d) with $L=100$ 
lattice sites and the potential amplitudes $V_0=4.6E_R\textrm{(c)};~9.9E_R\textrm{(d)}$,
thin (black) lines, at the temperatures $k_BT/\hbar\omega_{\rm box}=20$ (a); $10$ (b), 
$k_BT/J=0.5$ (c,d) are compared to Levy distributions, thick (red) lines. 
Here the frequency $\omega_{\rm box}=\hbar\pi^2/2ML_{\rm box}$ sets the energy 
scale, $\hbar \omega_0$, for a particle in a box. The momentum, $p$, is measured 
in units of $p_0=\hbar/L_{\rm box}$(a,b) and $p_0=\hbar k_L$ (c,d), in the case 
of a lattice potential and a box, respectively. The thin solid (blue) icons indicate the 
two corresponding types of confinement geometry.}
\label{fig5}
\end{figure}

So far we have been dealing with a TG gas on a lattice with an additional harmonic
potential (\ref{eq:hamiltonian}), which is the only confinement where the MD of
a TG gas was experimentally studied in detail \cite{bloch}. In the present
section, we demonstrate applicability of the Levy-spline approximation to the numerically
obtained MD of hard-core bosons at finite temperatures in various 1d confinements:
a sole harmonic trap, Fig.~\ref{fig4}, a box, Fig.~\ref{fig5}(a,b), and a sole
optical lattice with impenetrable boundaries, Fig.~\ref{fig5}(c,d).

A TG gas confined in a general potential, $V(x)$, is described by the sum of the 
single particle Hamiltonians,
\begin{equation}
\label{eq:continuum}
H=\sum_{i=1}^N \left(-\frac{\hbar^2}{2M}\frac{\partial^2}{\partial x_i^2}+V(x_i)\right)
\end{equation}
with the hard-core constrain on the bosonic many-particle wave function:
$\Psi(x_1,x_2,...,x_N)=0$ if $\vert x_i-x_j\vert<a$, where $x_i$ is the position
of $i$th particle, and $a$ is the 1d hard-core diameter. At low density, hard-bosons
can be approximated by impenetrable point size particles, so that $a=0$ \cite{tonk}.

The MD of $N$ hard-bosons in a box, $V_{\rm box}(x_i)=0$ if $0<x_i<L_{\rm box}$
and $V_{\rm box}=\infty$ otherwise, and in a harmonic confinement, $V_{\rm osc}=m\omega^2x_i^2/2$,
was obtained numerically within the canonical formalism with the help of efficient
method \cite{buljan}, which is detailed in Appendix~\ref{perez}. In the case of
a sole lattice potential with $L$ lattice sites, the MD was calculated within the
grand-canonical formalism by using the same algorithm that was employed in Sec.~IV
(see Appendix A).

It is known, that the MD of a TG gas in a homogeneous toroidal trap and in a harmonic
confinement (Fig.~\ref{fig4}) exhibits a power-law behavior, $p^{-4}$, for
$p\rightarrow\infty$ \cite{olshani}. We also found the same power-law behavior
for a TG gas in a box, see Fig.~\ref{fig5}(a,b). This in fact turns out to be a
general feature of a TG, which also persists at finite temperatures.
However, the power-law tail, $p^{-\alpha-1}$, of the Levy distribution with
exponent $\alpha$ cannot decay faster than $p^{-3}$ \cite{Levy}. Therefore, our Levy-spline
approximation is exclusively aimed to a finite momentum region $(0,p_{\rm c})$,
where the asymptotic $p^{-4}$ behavior has not yet developed. With Figs.~\ref{fig4}-\ref{fig5},
we demonstrate that the MD of a TG gas at finite temperatures in various 1d confinement
potentials (thin lines) can be perfectly approximated by Levy-splines (thick lines) over
significant momentum range.

While at zero temperature we detect a notable deviation of the Levy-spline from the
actual MD, with increasing the temperature the deviation becomes practically
negligible. The exponent of Levy-splines for a TG gas in the confinement potentials 
presented here depends, in general, on the confinement characteristics, on the gas temperature 
and on the particle number or the filling factor, $N/L$ (in the case of a lattice).
A more detailed analysis of this dependence will be addressed elsewhere.

\section{Conclusions}

We have presented a study of the finite-temperature momentum distribution
(MD) of Tonks-Girardeau (TG) gases confined in various 1d confinement potentials.
The MD of a TG gas on an optical lattice with a superimposed harmonic confinement
depends on the particle density, on the lattice depth and on the gas temperature.
We have shown that the tunable Levy distribution fits momentum profiles up
to one recoil momentum with high accuracy. This allows for calibration of
TG states with a unique scaling exponent. Thus, our approach completes the
attempts to quantify the finite-temperature MDs by using a power-law fitting
for the intermediate region $0.3 \leq |p/\hbar k_L| \leq 1$ \cite{bloch, pollet}.

We demonstrate that the MD of a TG gas confined in a 1d box, in a sole 1d harmonic
potential and on a 1d optical lattice with hard wall boundaries can be approximated
by the Levy distribution function (\ref{eq:Levy}) on a finite momentum region $(0,p_{\rm c})$,
where for a TG gas the universal power-law, $p^{-4}$, at large momenta has not yet
developed. We thus conjecture that Levy scaling of MD is generic feature of a 
thermalized TG gas.

We want to emphasize one aspect which we consider to be crucial for the understanding
of our approach. It is known that the Levy distribution has appeared as
the stable distribution, i.e. the ``attractor'' for normalized sums of independent
and identically-distributed random variables, with no finite mean values.
The latter has opened the door to anomalous \textit{statistics} \cite{Levy}.
In our approach, we employ the Levy distribution (\ref{eq:Levy}) merely as a
\textit{mathematical function}, which is a natural generalization of the standard
Gaussian function, thus leaving aside all the possible statistical interpretations
and speculations concerning physical processes responsible for the MD formation.
A good example for such ``applied'' approaches is the widespread use of the Gaussian
distribution, which describes the wave function of the ground state of the quantum
harmonic oscillator \cite{Liboff}, the Green's function solution of the deterministic
heat or diffusion equation \cite{heat}, and frequently results as a form function
in many different physical contexts.

While it is intuitively clear that it is the long-range correlations in the
system that cause the emergence of Levy distributions, the task to unravel the
inherent physical mechanism(s) yielding this anomalous distribution remains
a challenge. We think that the analysis of a reduced single-particle density
matrix in a spirit of the theory of random Levy matrices \cite{Levy_matrix}
may shed additional light on this intriguing issue.

\acknowledgments{We thank an anonymous referee for
useful suggestions and comments. This work was supported by the DFG through grant HA1517/31-1
and by the German Excellence Initiative ``Nanosystems Initiative Munich (NIM)''.}

\appendix

\section{Grand-canonical formalism}

Within the grand-canonical description, the number of bosons,
$N$, is a fluctuating quantity and the reduced single-particle density
matrix is defined as the trace over the Fock space,
\begin{equation}
\label{eq:gcdmatrix}
\rho_{n,l}=\frac{1}{Z}\textrm{Tr}\left\{\langle b_n^{\dagger}b_l\rangle e^{-\beta(H-\mu\sum_jn_j)}\right\},
\end{equation}
where the chemical potential $\mu$ is fixed to give the required number
of particles $N = \sum_l\rho_{l,l}$ in the system, $\beta = 1/k_BT$ with $k_B$
being the Boltzmann factor, and $Z$ is the grand-canonical partition function.
The latter coincides with that of non-interacting fermions, i.e.,
$Z = \prod_j[1+e^{-\beta(\epsilon_j-\mu)}]$, where $\epsilon_j$ stands
for the single-particle energy spectrum.

The trace over the the Fock space (\ref{eq:gcdmatrix}) can be evaluated exactly
\cite{grandcanon} by mapping the problem of hard-core bosons on that of spinless
fermions via the Jordan-Wigner transformation \cite{jordanwigner}. We next use the
expressions for the elements of the reduced single-particle density matrix
elaborated in \cite{grandcanon}, i.e.,
\begin{eqnarray}
\rho_{n,l} = \frac{1}{Z} &\left\{ \det \left[\mathbf{I} + (\mathbf{I}+\mathbf{A}(n,l))\mathbf{D}(n,l)\right]\right.  \nonumber \\
& \left. -\det\left[\mathbf{I} + \mathbf{D}(n,l)\right]\right\}\,,\mbox{for}\, n\neq l,
\end{eqnarray}
\begin{equation}
\mathbf{D}(n,l) = \mathbf{O}(l) \mathbf{U} e^{-(\mathbf{E}-\mu \mathbf{I})/k_B T}\mathbf{U}^\dagger \mathbf{O}(n);
\end{equation}
and
\begin{equation}
\rho_{n,n}=\left[\mathbf{U} \left(\mathbf{I}+e^{(\mathbf{E}-\mu \mathbf{I})/k_B T}\right)^{-1}
\mathbf{U}^\dagger \right]_{n,n},
\end{equation}
for the main diagonal elements. All operators entering above are the square matrices
$N\times N$ defined as follows: $\mathbf{I}$ denotes  the identity matrix,
$\mathbf{O}(n)_{i,j}=\{-\delta_{i,j}, i\leq n-1; \delta_{i,j},i>n-1\}$,
$\mathbf{A}(n,l)_{i,j}=\delta_{i,n}\delta_{j,l}$, $\mathbf{U}$ is the orthogonal
matrix of eigenvectors satisfying the eigenproblem, $H_1\mathbf{U}=\mathbf{U}\mathbf{E}$,
with single-particle version of Hamiltonian (\ref{eq:hamiltonian})
\begin{equation}
H_1=-J\sum_l (\vert l+1\rangle \langle l \vert + h.c.) + \nu\sum_l l^2\vert l\rangle \langle l \vert,
\end{equation}
and $\mathbf{E}$ is diagonal matrix of its eigenvalues, i.e.,
$\mathbf{E}_{i,j}=\varepsilon_i\delta_{i,j}$. Thus, to obtain the entire matrix
$\rho_{n,l}$ ($\rho_{l,n}=\rho_{n,l}^*$) one has to compute $N(N-1)/2$ determinants
of $N\times N$ matrices.

\section{Canonical formalism}

In the canonical formalism, the particle reservoir is absent,
meaning that one has to find the number, $\mathcal{N}_e$,
self-consistently from those many particle states, $|\Psi^m\rangle$,
together with their eigenenergies, $E_m$, that contribute to the thermal
superposition at a given temperature. The reduced single-particle density matrix
is then obtained as a sum of thermally weighted density matrices, $\rho^m$, 
evaluated for each from $\mathcal{N}_e$ eigenstates separately; i.e.,
\begin{equation}
\label{eq:cdmatrix}
\rho=\frac{1}{\mathcal{Z}}\sum_{m=1}^{\mathcal{N}_e}e^{-\beta E_m}\rho^m.
\end{equation}
The canonical partition function $\mathcal{Z}$ is expressed through the true many 
particle energy spectrum: $\mathcal{Z}=\sum_{m=1}^{\mathcal{N}_e}e^{-\beta E_m}$.

The energy spectrum of hard-core bosons is the same as that for spinless fermions:
$E_m=\sum_{n=1}^N \epsilon_{\alpha_n^m}$, where $\alpha_n^m$ denotes the numbers
of single-particle eigenlevels occupied in the $m$-th many-particle state.

\subsection{Hard-core bosons on a lattice}
\label{lattice}

In the case of a lattice confinement, $\rho^m\equiv\rho_{n,l}^m=\langle\Psi^m| b_n^{\dagger}b_l|\Psi^m\rangle$.
The corresponding many-particle eigenstates can be represented as
\begin{equation}
|\Psi^m\rangle = \prod_{\alpha_n^m} \sum_l \mathbf{U}_{l,\alpha_n^m} b_l^\dagger\vert 0\rangle, \, n=1,\dots,N
\end{equation}
where $\mathbf{U}$ is complete orthogonal set of single-particle eigenvectors
(see Appendix A).

Each contribution, $\rho_{i,j}^m$, is related to the Green function,
$G_{i,j}^m = \langle\Psi^m|b_ib_j^{\dagger}|\Psi^m\rangle$:
\begin{equation}
\rho_{i,j}^m = G_{i,j}^m +\delta_{i,j}(1-2G_{i,j}^m).
\end{equation}
Using the Jordan-Wigner transformation \cite{jordanwigner}, the bosonic Green
function can be rewritten as scaler product of two fermionic wave-functions,
and subsequently as a determinant of matrix product \cite{canon}:
\begin{equation}
G_{i,j}^m = {\rm det}\left[(\mathbf{P}_m(i))^\dagger\mathbf{P}_m(j)\right],
\end{equation}
where the matrix $\mathbf{P}_m(i)_{l,n}$ has $N+1$ columns
\begin{equation}
\mathbf{P}_m(i)_{l,n} = \left\{
\begin{array}{cl}
-\mathbf{U}_{l,\alpha_n^m},\, & \mbox{for} \, l\leqslant i-1, n=1,\dots,N; \\
 \mathbf{U}_{l,\alpha_n^m},\, & \mbox{for} \, l> i-1, n=1,\dots,N; \\
 \delta_{i,l},\, & \mbox{for} \, n=N+1. \\
\end{array}
\right.
\end{equation}
In comparison to the grand-canonical approach, the number of operations needed to
obtain the entire matrix $\rho_{i,j}$ is a factor of $\mathcal{N}_e$ larger, and
is growing with increasing temperature.

\subsection{Hard-core bosons in arbitrary confinement potential}
\label{perez}

Above we detailed the canonical formalism for hard-core bosons on a lattice.
Here, instead, we elaborate on a general case with arbitrary confinement, $V(x)$. 
In this case, the momentum distribution, $n(p)$, can be obtained from the reduced 
single-particle density matrix in the continuum: $n(p)=(2\pi)^{-1}\int dxdye^{-ik(x-y)}\rho(x,y)$,
where $\rho(x,y)$ is given by the thermal superposition (\ref{eq:cdmatrix}) with
$\rho^m\equiv\rho^m(x,y)$ defined below.

To calculate the reduced single-particle density matrix, $\rho^m(x,y)$, in the continuum
we make use of efficient method \cite{buljan}, which represents it in terms of the single 
particle states:
\begin{equation}
\label{eq:perez}
\rho^m(x,y)=\sum_{i,j=1}^N \psi_{\alpha_i^m}^*(x)A^m(x,y)_{i,j}\psi_{\alpha_j^m}(y),\\
\end{equation}
where $N\times N$ matrix $\mathbf{A}^m(x,y)$ associated to the $m$-th many-particle state is
\begin{equation}
\label{eq:perez_sup_1}
A^m(x,y)_{i,j}=\left(-1\right)^{i+j}\det \mathbf{Q}_{m}(i,j).
\end{equation}
$\mathbf{Q}_{m}(i,j)$ is a minor of matrix $\mathbf{Q}_m$ obtained by crossing $i$-th row and
$j$-th column, and matrix $\mathbf{Q}_m$ itself is given as
\begin{equation}
\label{eq:perez_sup_2}
Q_m(x,y)_{i,j} = \delta_{i,j}-2\int_x^ydx'\psi_{\alpha_i^m}^*(x')\psi_{\alpha_j^m}(x'),
\end{equation}
where $\psi_n(x)$ are the eigenfunctions of single particle eigenproblem in a given
trapping potential $V(x)$. In (\ref{eq:perez_sup_2}), it is assumed that $x\geq y$,
while for $x\leq y$: $\rho^m(x,y)=\rho^m(y,x)^*$. The set of $\alpha_i^m$, as before in 
Appendix \ref{lattice}, denotes the numbers of single-particle eigenlevels occupied in the $m$-th many-particle state.
Additionally, whenever $\det\mathbf{Q}_{m} \neq 0$, (\ref{eq:perez_sup_1}) can be represented as
$\mathbf{A}^m = \left(\mathbf{Q}_m^{-1}\right)^T \det{\mathbf{Q}_m}$ \cite{buljan}, 
which is more efficient when implemented numerically.

For a box and for a harmonic confinement, $\psi_n(x)=\sqrt{2/L_{\rm box}}\sin(n\pi x/L_{\rm box})$ and
$\psi_n(x)=\sqrt{1/2^n n!}\left(m\omega_0/\pi\hbar\right)^{1/4}\exp\left(-m\omega_0x^2/2\hbar\right)H_n(\sqrt{m\omega_0/\hbar}x)$,
respectively, where $H_n(y)$ are the Hermite polynomials. For these confinements, the integral in
(\ref{eq:perez_sup_2}), and the density matrices $\rho^m(x,y)$ can be obtained analytically
for a small number of particles with the help of symbolic computational routines. However,
expanding out $m$ times $N^2$ determinants of $(N-1)\times(N-1)$ matrices for $N\gtrsim 5$ quickly
becomes unwieldy. Therefore, in Sec.~V, the reduced density matrices $\rho^m(x,y)$
were calculated on a numerical grid. The grid step was sufficiently small to ensure a smooth
representation of all single particle wave-functions, $\psi_m(x)$, participating in excited
many-particle states at a given temperature.


\begin{thebibliography}{1000}

\bibitem{thoulfetter} D. J. Thouless,
\textit{The Quantum Mechanics of Many-body Systems}
(Academic Press, New York, 1972);
A. L. Fetter and J. D. Walecka, \textit{Quantum Theory of Many-Particle Systems}
(Dover, New York, 2003).

\bibitem{metcalf} H. J. Metcalf and P. van der Straten,
\textit{Laser Cooling and Trapping} (Springer, New York, 1999).

\bibitem{ober} O. Morsch and M. Oberthaler,
Rev. Mod. Phys. \textbf{78}, 179 (2006).

\bibitem{zwerger} I. Bloch, J. Dalibard, and W. Zwerger,
Rev. Mod. Phys. \textbf{80}, 885 (2008).

\bibitem{tonk} M. Girardeau, J. Math. Phys. \textbf{1},  516 (1960).

\bibitem{bloch} B. Paredes {\it et al.}, Nature \textbf{429},
277 (2004).

\bibitem{kinoshita1} T. Kinoshita, T. Wenger, and D. S. Weiss,
Science \textbf{305}, 1125 (2004).

\bibitem{olshani} M. Olshanii and V. Dunjko,
Phys. Rev. Lett. \textbf{91}, 090401 (2003).

\bibitem{papen} T. Papenbrock, Phys. Rev. A \textbf{67}, 041601(R) (2003).

\bibitem{minguzzi} A. Minguzzi, P. Vignolo, and M. P. Tosi,
Phys. Lett. A \textbf{294}, 222 (2002).

\bibitem{pollet} L. Pollet, S. M. A. Rombouts, and P. J. H. Denteneer,
Phys. Rev. Lett. \textbf{93}, 210401 (2004).

\bibitem{Levy} W. Feller, \textit{An Introduction to Probability Theory
and Its Applications} (John Wiley and Sons, New York, 1970), Vol.2.

\bibitem{Klafter} M. F. Shlesinger, G. M. Zaslavsky, and J. Klafter,
Nature \textbf{363}, 31 (1993); A. Blumen, G. Zumofen, and J. Klafter,
Phys. Rev. A \textbf{40}, 3964 (1989).

\bibitem{sub} F. Bardou, J.-P. Bouchaud, A. Aspect and C.
Cohen-Tannoudji, \textit{Levy Statistics and Laser Cooling}
(Cambridge Univ. Press, Cambridge 2000).

\bibitem{econ} R. N. Mantegna and H. E. Stanley,
Nature \textbf{376}, 46 (1995).

\bibitem{barkai} E. Barkai, R. Silbey, and G. Zumofen,
Phys. Rev. Lett. \textbf{84}, 5339 (2000).

\bibitem{network} R. Segev \textit{et al.},
Phys. Rev. Lett. \textbf{88}, 118102 (2002).

\bibitem{heart} C.-K. Peng  \textit{et al.},
Phys. Rev. Lett. \textbf{70}, 1343 (1993).

\bibitem{grandcanon} M. Rigol,
Phys. Rev. A \textbf{72}, 063607 (2005).

\bibitem{canon} M. Rigol and A. Muramatsu,
Phys. Rev. A \textbf{70}, 031603(R) (2004); {\it ibid.}, \textbf{72}, 013604 (2005).

\bibitem{BHH} D. Jaksch, C. Bruder, J. I. Cirac, C. W. Gardiner, and
P. Zoller, Phys. Rev. Lett. \textbf{81}, 3108 (1998).

\bibitem{pendulum} The critical density $n_s$ is found via mapping the single
particle version of (\ref{eq:hamiltonian}) onto the quantum pendulum \cite{avp}.

\bibitem{avp} A. V. Ponomarev and A. R. Kolovsky,
Laser Phys. \textbf{16}, 367 (2006).

\bibitem{superstat} C. Beck and E. G. D. Cohen,
Physica A \textbf{322}, 267 (2003).

\bibitem{buljan} R. Pezer and H. Buljan,
Phys. Rev. Lett. \textbf{98}, 240403 (2007).

\bibitem{Liboff} R. L. Liboff,
\textit{Introductory Quantum Mechanics} (Addison-Wesley, London, 2002).

\bibitem{heat} G. Barton,
\textit{Elements of Green's functions and Propagation} (Clarendon press, Oxford, 1989).

\bibitem{Levy_matrix} P. Cizeau and J. P. Bouchaud,
Phys. Rev. E \textbf{50}, 1810 (1994).

\bibitem{jordanwigner} P. Jordan and E. Wigner,
Z. Phys. \textbf{47}, 631 (1928).


\end{thebibliography}
\end{document}